\documentclass[preprintnumbers,amsmath,amssymb,nofootinbib]{revtex4}

\usepackage{graphicx}
\usepackage{dcolumn}
\usepackage{bm}
\usepackage{color}

\newcommand\ee{\end{equation}}
\newcommand\be{\begin{equation}}
\newcommand\eea{\end{eqnarray}}
\newcommand\bea{\begin{eqnarray}}

\begin{document}


\title{Radiation and the classical double copy for color charges}

\author{Walter D. Goldberger}
\affiliation{Physics Department, Yale University, New Haven, CT 06520, USA}
\author{Alexander K. Ridgway}
\affiliation{Walter Burke Institute for Theoretical Physics, California Institute of Technology, Pasadena, CA 91125, USA}

\date{\today}

\begin{abstract}

We construct perturbative classical solutions of the Yang-Mills equations coupled to dynamical point particles carrying color charge.   By applying a set of color to kinematics replacement rules first introduced by Bern, Carrasco and Johansson (BCJ), these are shown to generate solutions of $d$-dimensional dilaton gravity, which we also explicitly construct.       Agreement between the gravity result and the gauge theory double copy implies a correspondence between  non-Abelian particles and gravitating sources with dilaton charge.   When the color sources are highly relativistic, dilaton exchange decouples, and the solutions we obtain match those of pure gravity.   We comment on possible implications of our findings to the calculation of gravitational waveforms in astrophysical black hole collisions, directly from computationally simpler gluon radiation in Yang-Mills theory.

\end{abstract}

\maketitle

\section{Introduction}

The many structural similarities between Yang-Mills theory and general relativity hint at possible relations between their observables and dynamics.    At the perturbative level, a correspondence between $S$-matrix elements in gauge theory and gravity was discovered in recent years by Bern, Carrasco and Johansson (BCJ)~\cite{BCJ}.     Their result builds on earlier work of~\cite{KLT} which found certain squaring identities that relate tree-level open and closed string amplitudes.    In particular, the BCJ relations contain as a special case the KLT identities of~\cite{KLT} in the limit of large string tension, in which the infinite tower of massive string modes decouples.

The BCJ relations state that once written in a specific form, Yang-Mills amplitudes, in any spacetime dimension $d$, can be mapped onto their gravity counterparts by applying a simple set of well-defined color-to-kinematics replacement rules.     As in the KLT case, the double copy of gauge theories with or without matter is not pure general relativity.   Rather, as expected from counting degrees of freedom, the gravitational theory generally contains other massless fields in addition to the graviton.   For example, in pure Yang-Mills theory, the double copy is a theory whose massless states consist of a graviton, a scalar field $\phi$ (dilaton), and a 2-form gauge field $B_{\mu\nu}$.   The BCJ double copy has been verified for amplitudes in a wide number of field theories, containing varying amounts of supersymmetry, at the multi-loop level.    In field theory, these relations have been explicitly derived~\cite{Bern:2010yg} for $n$-point tree-level amplitudes, using modern amplitude techniques.   However, the microscopic origin of these relations, either in field theory or string theory, is still unknown.   See~\cite{Carrasco:2015iwa} for a review and more extensive references to the literature.

Aside from their inherent theoretical interest, the BCJ relations are of practical significance, as they reduce the computational complexity of perturbative gravity to the relatively more manageable gauge theory Feynman rules.     It is therefore natural to consider if similar relations can arise for  observables beyond the $S$-matrix.   For example, is it possible to predict the classical gravitational radiation field generated by a system of merging black holes from the analogous solution in gauge theory coupled to color sources?    This would have potential astrophysical applications, to the calculation of templates for gravitational wave detectors such as LIGO\footnote{An effective field theory of gravitons whose Feynman diagrams compute gravitational radiation from merging black holes was first introduced in~\cite{GnR}.  The possibility that gravity is the square of gauge theory, together with on-shell methods, can be used to simplify gravitational wave calculations was discussed by~\cite{iraandduff}.}.

The existence of non-perturbative double copy relations between classical solutions in gauge theory and gravity was raised first in~\cite{Monteiro:2014cda}.     They proposed a correspondence between solutions in the Abelian sector~\cite{jackiwrev} of Yang-Mills theory and Kerr-Schild spacetimes in general relativity.   The gauge theory configurations consist of solutions of the form $A_\mu^a(x) = c^a \chi(x) k_\mu(x)$ up to gauge (with $c_a$ constant and $k^\mu$ null), for which the Yang-Mills equations become linear.     These solutions have close counterparts in general relativity consisting of metrics that up to gauge are of the Kerr-Schild form
\be
g_{\mu\nu} = \eta_{\mu\nu} + m \chi(x) k_\mu k_\nu.
\ee 
An important example is the non-Abelian $1/r^{d-3}$ Coulomb field of a point color charge $c^a$,  corresponding to the $d$-dimensional Schwarzschild metric with mass $m$, which is indeed in the Kerr-Schild class.   Further examples can be found in~\cite{Monteiro:2014cda,Luna:2015paa,Luna:2016due,MarkandAlec}.    See~\cite{Chu:2016ngc} for related discussion.

The Kerr-Schild correspondence $c^a\rightarrow m$ of~\cite{Monteiro:2014cda} is a natural extension of the BCJ relations to the non-perturbative regime\footnote{See~\cite{CNN1,CNN2} for examples of non-perturbative double copy structure in supersymmetric theories.}.   However, it is unclear how to use this approach to obtain more generic solutions in gravity which deviate from the algebraically special Kerr-Schild form.    In this paper, we instead focus on perturbative solutions, and find evidence that BCJ-type relations exist between classical solutions of Yang-Mills theory and those in a gravitational double copy containing scalar and graviton degrees of freedom.     On the Yang-Mills side, the setup consists of several initially well-separated color charges, interacting through gluon exchange.   These sources are not treated as fixed.  Instead, they evolve self-consistently in the classical gluon field they themselves generate.    We construct the long distance radiation gluon field of this system, expressed in terms of momentum space integrals.    By applying BCJ-type color-to-kinematics substitutions at the level of the integrand, these are shown to precisely match, at leading perturbative order, the asymptotic radiation fields in a theory containing gravity and a massless scalar (dilaton) coupled to massive point particles.

More explicitly, in sec.~\ref{sec:YM}, we construct the Yang-Mills radiation field corresponding to a set of color charges coming in from spatial infinity, in terms of the initial momenta $p^\mu$ and initial charges $c^a$ transforming in the adjoint representation of the gauge group.    Our calculation is done in general spacetime dimension $d$, generalizing known $d=4$ static,~\cite{sikivie,jackiw} and radiating~\cite{gyulassi,Kovchegov:1997ke} two-particle solutions of the classical Yang-Mills equations.   Exact classical solutions of classical $SU(N\rightarrow\infty)$ QCD coupled to color sources were constructed in~\cite{Kosyakov:1998qi}.  Even though the observables we obtain are classical, they are related to the quantum mechanical on-shell gluon emission amplitude sourced by semi-classical non-Abelian particles.  

In sec.~\ref{sec:grav}, we consider analogous solutions in the graviton-dilaton system coupled to massive (non-spinning) point-particles.    In addition to computing the asymptotic radiation gravitational and scalar fields sourced by these particles, we compute the gravitational analog of the Wilson line which evolves each particle's momentum from initial to late times.     We find in sec.~\ref{sec:doublecopy} a double-copy relation between these observables and the corresponding quantities in the classical gauge theory.    The specific relation takes the initial color charges $c^a$ and replaces them by a second copy of the initial momenta $p^\mu$, and maps the Lie algebra structure constants $f^{abc}$ to a second copy of the kinematic part of the three-gluon Feynman rule (the four-gluon interaction does not yet show up at the order in perturbation theory that we consider here).     These substitutions transform the asymptotic gluon field detected by far away observers to a double-copy radiation amplitude that encodes emission in both graviton and dilaton channels.  

For the double-copy relations discussed in sec.~\ref{sec:doublecopy} to work, it is essential to choose the dilaton couplings to the point particles in such a way that the asymptotic scalar and radiation fields have, at the level of the integrand, no explicit dependence on the dimensionality $d$.   The specific form of the interactions is motivated by observations made in~\cite{berngrant}, and require the point-particles to have scalar coupling strengths proportional to mass.    Thus, at least perturbatively, the correspondence between gauge and gravity solutions implies that the non-Abelian Coulomb field of a color charge maps onto a configuration which on the gravity side has non-vanishing scalar field profile.   In the case of a strictly massless source, or in the limit $d\rightarrow\infty$ of large spacetime dimensions\footnote{The limit $d\rightarrow\infty$  of perturbative quantum gravity was discussed in refs.~\cite{Strominger:1981jg,BjerrumBohr:2003zd} and for solutions of classical general relativity in~\cite{Emparan:2013moa}.}, the scalar field configuration vanishes and the resulting gravitational field is of Kerr-Schild form, in agreement with the ideas of~\cite{Monteiro:2014cda}.   However, away from these limits the perturbative double copy of a charged non-Abelian particle is not the Schwarzschild solution and, as we explicitly check in sec.~\ref{sec:ptgmunu}, the gravitational field is coordinate inequivalent to Kerr-Schild form starting at second order in perturbation theory.  This suggests that the duality proposed by~\cite{Monteiro:2014cda} may only hold in certain domains.

More generally, as we discuss in sec.~\ref{sec:massless}, for solutions corresponding to several massless interacting particles, the dilaton can be consistently decoupled to all orders in perturbation theory and the double copy reduces to pure general relativity.    This observation may be useful for computing gravitational wave templates from merging black holes directly from classical gauge theory, albeit in a limit in which the sources are highly boosted yet still within the perturbative regime.    We also comment in sec.~\ref{sec:conc} on the inclusion of worldline spin degrees of freedom and the related question of the role played by the two-form gauge field $B_{\mu\nu}$ in the classical double copy, as well as other open questions raised by the results presented in this paper.

\section{Classical Yang-Mills solutions}
\label{sec:YM}

\subsection{Equations of motion and classical observables}

We consider solutions of classical Yang-Mills theory in $d$-spacetime dimensions coupled to point particle color charges.   By definition, these are objects localized on a worldline $x^\mu(\tau)$ that carry a color charge degree of freedom $c^a(\tau)$ transforming in the adjoint representation of the gauge group~\footnote{Our conventions are $D_\mu =\partial_\mu + i g A^a_\mu T^a$, $[T^a,T^b]=if^{abc} T^c$.  The generators in the adjoint representation are $\left(T_{\mbox{\tiny{adj}}}^a\right)^b{}_c = - i f^{abc}$.}.   The equations of motion are
\begin{equation}
\label{eq:YM}
D_\nu F^{\nu\mu}_a(x) = g J_a^\mu(x),
\end{equation}
where the color current is given by 
\be
\label{eq:ccurrent}
J^\mu_a(x)=\sum_\alpha \int d\tau c_{\alpha}^a(\tau) v^\mu_\alpha(\tau) \delta^d(x-x_\alpha(\tau)).
\ee
Here $\alpha$ is a label that distinguishes the different point masses and $v_\alpha^\mu=dx_\alpha^\mu/d\tau$ is their velocity.  Current conservation, $D_\mu J^\mu_a=0$ gives rise to the parallel transport equation $v\cdot D c^a =0$ for each charge, i.e.
\be
\label{eq:cpt}
{d c^a\over d\tau} = g f^{abc} v^\mu A^b_\mu(x(\tau)) c^c(\tau),
\ee
with well known solution $c_\alpha^a(\tau) = {W_\alpha}^a{}_b(\tau) c_\alpha^b(-\infty)$ in terms of the adjoint representation Wilson line along the trajectory $x^\mu_\alpha(\tau)$,
\be
{W_\alpha}^a{}_b(\tau) = \left[P \exp\left\{-ig\int^\tau_{-\infty} dx_\alpha^\mu  A_\mu\cdot T_{\mbox{\tiny{adj}}}\right\}\right]^a{}_b.
\ee
The orbital motion follows from the conservation of total energy-momentum $\partial_\mu (T^{\mu\nu}_{YM} + T^{\mu\nu}_{pp})=0$, with
\be
T_{pp}^{\mu\nu}(x) = \sum_{\alpha} m_{\alpha} \int d\tau v^\mu_\alpha(\tau) v^\nu_\alpha(\tau) \delta^d(x-x_\alpha(\tau))
\ee
which implies that each particle obeys the Lorentz force law
\be
\label{eq:lorentz}
{d p^\mu\over d\tau} = m {d^2 x^\mu\over d\tau^2} = g  c^a F_a^{\mu}{}_{\nu}  v^\nu.
\ee

The classical equations of motion Eq.~(\ref{eq:cpt}), Eq.~(\ref{eq:lorentz}) were first obtained in~\cite{Wong} by taking a limit of the Dirac equation.  There are several physically distinct Lagrangian realizations.   For instance, one can take as a specific model the Lagrangian~\cite{bala}, 
\be
S_{pp} = -m \int d\tau + \int d\tau \psi^\dagger iv\cdot D \psi,
\ee 
where $\psi(\tau)$ is a variable transforming linearly at $x^\mu(\tau)$ under the gauge group, and $c_a=\psi^\dagger T^a \psi$.    See also~\cite{nll} for a different Lagrangian formulation with non-linearly realized gauge symmetry on the worldline.   For the purposes of this paper, it is sufficient to work directly in terms of Eqs.~(\ref{eq:cpt}), (\ref{eq:lorentz}), so whatever results we obtain will hold independently of any particular action formulation.  

\begin{figure}
\centering
\includegraphics[scale=0.3]{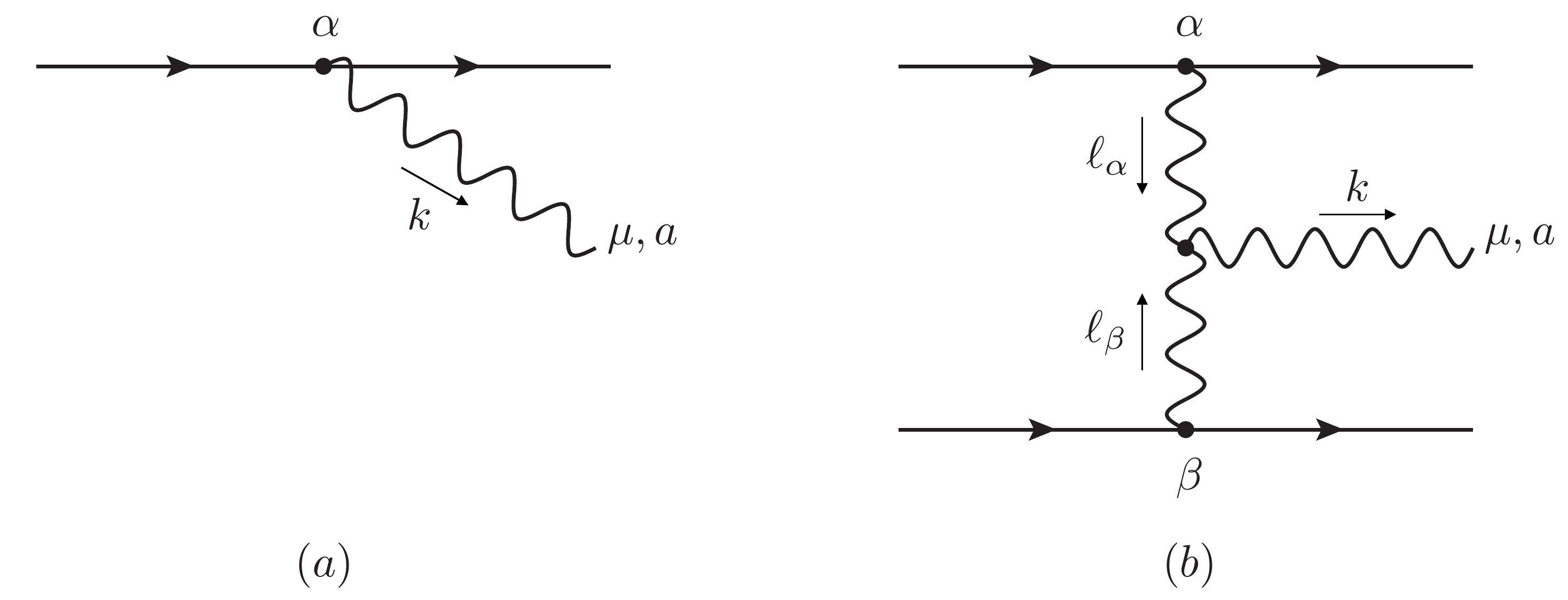}
\caption{Leading order Feynman diagrams for the perturbative expansion of ${\tilde J}^\mu_a(k)$.\label{fig:gluon1pt}}    
\end{figure}

The main object of interest is the self-consistent classical field $\langle A^a_\mu\rangle(x)$ generated by a collection of point color charges that evolve according to the equations of motion Eqs.~(\ref{eq:YM}),~(\ref{eq:cpt}),~(\ref{eq:lorentz}).    As long as the particles remain well separated, we can determine $\langle A^a_\mu\rangle(x)$ perturbatively as an expansion in powers of the gauge coupling $g$.   We find it convenient to work in the gauge $\partial_\mu A^\mu_a=0$, in which case the Yang-Mills equation can be re-written as~\cite{jackiwrev}
\be
\label{eq:ccurrentdef}
\Box A^\mu_a = g {\tilde J}^\mu_a,
\ee
where the gauge-dependent current ${\tilde J}^\mu_a$ is defined as
\begin{eqnarray}
\nonumber
{\tilde J}^\mu_a &=& J^\mu_a +  f^{abc} A^b_\nu(\partial^\nu A_c^\mu - F_c^{\mu\nu}),\\
 \partial_\mu {\tilde J}^\mu_a &=& 0.
\end{eqnarray}
Formally, Eq.~(\ref{eq:ccurrentdef}) can be solved iteratively.   Once the solution at given order in $g$ is found, it is fed back in to get the field at the next order in perturbation theory.  Equivalently, it is useful to to adopt a diagrammatic approach, where the classical solution $\langle A_\mu\rangle(x)$ to Eq.~(\ref{eq:ccurrentdef}) is calculated as a sum of Feynman diagrams of the form shown in Fig~\ref{fig:gluon1pt}.    These diagrams are computed using standard momentum space Feynman rules, with insertions of the (Fourier transformed) current Eq.~(\ref{eq:ccurrent}).    At the classical level, in order to preserve causality, it is necessary to use a retarded, or ``in-in" $i\epsilon$ prescription for the gluon propagator\footnote{This can be justified by interpreting $\langle A_\mu\rangle(x)$ as the tree level part of the in-in one-point correlation function $\langle in|A^a_\mu(x) |in\rangle$ in quantum field theory coupled to classical particles.}.   This is in contrast to the standard Feynman boundary condition that must be used to compute $S$-matrix elements between asymptotic in/out states.   In this paper, it is implicit that propagators obey retarded boundary conditions, i.e. $1/k^2 = 1/[(k^0+i\epsilon)^2 - {\vec k}^2]$.

Once the classical solution $\langle A_\mu\rangle(x)$ is known to a given order in perturbation theory, it can be used to compute all the physical observables of this system.    Here, we focus on observables measured by asymptotic observers at spatial infinity, $r=|{\vec x}|\rightarrow\infty$, which are directly related to the momentum space current ${\tilde J}^\mu_a(k)=\int d^d x e^{ik\cdot x} {\tilde J}^\mu_a(x)$ evaluated for on-shell momentum $k^2=0$.   For example in $d=4$ spacetime dimensions the asymptotic field at $r\rightarrow\infty$ and retarded time $t$ is
\be
\lim_{r\rightarrow\infty}\langle  A^a_\mu\rangle(x)  = {g\over 4\pi r}\int {d\omega\over 2\pi} e^{-i\omega t} {\tilde J}^\mu_a(k)
\ee
with $k^\mu = (\omega,{\vec k})=\omega(1,{\vec x}/r)$.    Similar results exist also in $d$ dimensions.    Thus the on-shell current  ${\tilde J}^\mu_a(k)$ directly measures the flux of energy-momentum, color, and angular momentum radiated out to infinity by the system of point charges.   In particular, the total energy-momentum radiated out to $r\rightarrow\infty$ in a fixed polarization channel $\epsilon^a_\mu(k)$ is ($\int_k=\int d^d k/(2\pi)^d$)
\be
\Delta P^\mu = \int_{k} (2\pi)\theta(k^0)\delta(k^2)  \left|{\cal A}^a(k)\right|^2 k^\mu,
\ee
where we have defined the on-shell radiation amplitude ${\cal A}^a(k) = \left.\epsilon_\mu^a(k) \left[g {\tilde J}^\mu_a(k)\right]\right|_{k^2=0}$, with the polarization vector obeying $k\cdot \epsilon^a(k)=0$, $\epsilon^a(k)\cdot\epsilon^b(k)^*=-\delta_{ab}$.   Similar expressions hold for other conserved quantities (angular momentum and color charge).

\subsection{Perturbative solutions}

We consider a setup consisting of several particles $\alpha=1,\ldots, N$ coming in from infinity at $\tau\rightarrow-\infty$, with initial data
\be
c_\alpha^a(\tau\rightarrow-\infty) = c_\alpha^a,
\ee
\be
x^\mu_\alpha(\tau\rightarrow-\infty) = b^\mu_\alpha + v^\mu_{\alpha} \tau,
\ee
where $b^\mu_\alpha$ are spacelike (the impact parameters are $b_{\alpha\beta}^\mu = b_\alpha^\mu-b_\beta^\mu$) and $v_\alpha^2=1$.    As the particles come close and interact, they scatter and emit classical radiation encoded in $\langle A_\mu\rangle(x)$.   We assume that the charges always remain sufficiently far away so that the particles' trajectory changes are small and perturbation theory is valid.     In this case, the classical radiation field can be computed formally in powers of $g$, in terms of the diagrams shown in Fig.~\ref{fig:gluon1pt}.   The precise form of the dimensionless small parameter that governs this expansion depends on the kinematics, and can be determined by estimating the dependence of higher order diagrams not shown in Fig.~\ref{fig:gluon1pt} on the kinematic variables.  E.g., for particles of comparable mass and energy $E\gtrsim m$, one finds the expansion parameter is
\be
\label{eq:gteps}
\epsilon = g^2 {\Gamma\left({d-3\over 2}\right)\over (4\pi)^{{d-2\over 2}}} {c_\alpha\cdot c_\beta \over E b_{\alpha\beta}^{d-3}}\ll 1.
\ee
Since we construct our solutions only at the level of the integrand, we do not need to make any assumptions about the size of the radiation frequency scale $\omega$ relative to the typical impact parameter $b$.  Note that even though the diagrams in Fig.~\ref{fig:gluon1pt} scale like definite powers of the gauge coupling, they each contain in general an infinite series of terms suppressed by powers of $\epsilon\ll 1$.    It should be possible to construct a classical effective field theory whose Feynman diagrams scale as definite powers of $\epsilon$  and/or $\omega b\ll 1$ along the lines of ref.~\cite{GnR}.    However, since in this paper we are only interested in the leading order radiation field, we do not find it necessary to introduce manifest power counting in $\epsilon\ll 1$.   Rather, we truncate $\mbox{Fig.}~\ref{fig:gluon1pt}(a)= \epsilon^0 + \epsilon^1+\cdots$ and $\mbox{Fig.}~\ref{fig:gluon1pt}(b)=\epsilon^1 + \cdots$.

At intermediate times, the particle trajectories in physical and color space are given by
\be
x_\alpha^\mu(\tau) = b^\mu_\alpha + v^\mu_\alpha\tau + z^\mu_\alpha(\tau),
\ee
\be
c_\alpha^a(\tau) = c^a_\alpha + {\bar c}^a_\alpha(\tau),
\ee
where the deflections $z^\mu_\alpha(\tau)$, ${\bar c}^a(\tau)$ contain terms of order $g^2$ (i.e $\epsilon^1$) and higher.    To compute these deflections, we first need the classical field at leading order, which is given by the diagram in Fig.~\ref{fig:gluon1pt}(a) with static sources $z^\mu={\bar c}^a=0$.    In Feynman gauge, this is 
\be
\label{eq:LOA}
\langle A_a^\mu \rangle(x) = \sum_\alpha\int_\ell -{i\over \ell^2} e^{-i\ell\cdot x} (-i g)\int d\tau c_\alpha^a v_\alpha^\mu e^{i\ell\cdot(b_\alpha + v_\alpha\tau)} = -g\sum_\alpha\int_\ell (2\pi)\delta(\ell\cdot v_\alpha) {e^{-i\ell\cdot(x-b_\alpha)}\over\ell^2} v_\alpha^\mu c^a_\alpha.
\ee
At this order in perturbation theory, the current is ${\tilde J}^a_\mu(k) = g\sum_\alpha e^{ik\cdot b_\alpha} (2\pi)\delta(k\cdot v_\alpha) v^\mu_\alpha c^a_\alpha,$ which vanishes on-shell for timelike $v^\mu_\alpha$, since static color charges do not radiate.   Inserting this result into the equations for $x^\mu(\tau)$ and $c^a(\tau)$,  we can now obtain the leading order deflections
\be
\label{eq:cdotg2}
{d{\bar c}^a_\alpha(\tau)\over d\tau} = - g^2 \sum_{\beta\neq\alpha}(v_\alpha\cdot v_\beta) f^{abc} c_\beta^b c_\alpha^c \int_\ell (2\pi) \delta(\ell\cdot v_\beta) {e^{i\ell\cdot (b_{\alpha\beta}+ v_\alpha\tau)}\over\ell^2} ,
\ee
and
\be
\label{eq:zddotg2}
m_\alpha {d^2 z^\mu_\alpha\over d\tau^2} =  -ig^2 \sum_{\beta\neq\alpha}(c_\alpha\cdot c_\beta) \int_\ell (2\pi) \delta(\ell\cdot v_\beta) {e^{i\ell\cdot (b_{\alpha\beta} + v_\alpha\tau)}\over\ell^2} \left[(v_\alpha\cdot v_\beta) \ell^\mu - (v_\alpha\cdot \ell) v_\beta^\mu\right].
\ee 
In particular, from Eq.~(\ref{eq:cdotg2}) we can read off the Wilson line matrix at order  $g^2$
\be
[\ln W_\alpha(\tau\rightarrow\infty)]^a{}_b = g^2 \sum_{\beta\neq\alpha} \int_\ell (2\pi)\delta(\ell\cdot v_\alpha) (2\pi)\delta(\ell\cdot v_\beta) {e^{i\ell\cdot b_{\alpha\beta}}\over \ell^2} (v_\alpha\cdot v_\beta) f^{abc} c^c_\beta.
\ee
For later comparison to gravity, it is useful to insert a spurious integral over a new momentum variable $\ell_\alpha$ and re-label $\ell\rightarrow -\ell_\beta$,
\be
\label{eq:ymw}
[\ln W_\alpha(\tau\rightarrow\infty)]^a{}_b= g^2\sum_{\beta\neq\alpha} \int_{\ell_\alpha,\ell_\beta} \mu_{\alpha,\beta}(k=0)\, 
\ell_\alpha^2 (v_\alpha\cdot v_\beta) f^{abc} c^c_\beta,
\ee
where we have introduced the notation
\be
\mu_{\alpha,\beta}(k) = 
\left[(2\pi)\delta(v_\alpha\cdot\ell_\alpha)  {e^{i\ell_\alpha\cdot b_\alpha}\over\ell^2_\alpha}\right] \left[(2\pi)\delta(v_\beta\cdot\ell_\beta)  {e^{i\ell_\beta\cdot b_\beta}\over\ell^2_\beta}\right] (2\pi)^d \delta^d(k-\ell_\alpha-\ell_\beta).
\ee
The momentum integrals appearing in these expressions can be done by standard Schwinger parameter methods, but we will not need those in what follows.

The order $g^2$ (or $\epsilon^1$) correction to the current ${\tilde J}^\mu_a(k)$ field has two contributions.   One is from the diagram in Fig.~\ref{fig:gluon1pt}(a), taking into account the order $g^2$ corrections to the equations of motion in Eqs.~(\ref{eq:cdotg2}),~(\ref{eq:zddotg2}).   This gives rise to the following contribution to the conserved color current ${\tilde J}^\mu_a(k)$ defined in Eq.~(\ref{eq:ccurrentdef}):
\be
\left.\mbox{Fig.}~\ref{fig:gluon1pt}(a)\right|_{{\cal O}(g^2)} = \sum_\alpha e^{i k\cdot b_\alpha}\left[-i (k\cdot v_\alpha) c^a_\alpha \left(z^\mu_\alpha(\omega) - {k\cdot z_\alpha\over k\cdot v_\alpha} v^\mu_{\alpha}\right) + {\bar c}^a_\alpha(\omega) v^\mu_\alpha\right]_{\omega=k\cdot v_\alpha},
\ee
where $z^\mu(\omega)=\int d\tau e^{i\omega\tau} z^\mu(\tau)$, ${\bar c}^a(\omega)=\int d\tau e^{i\omega\tau} {\bar c}^a(\tau)$ are the frequency-space displacements.   In this equation and in what follows, the $i\epsilon$ prescription $1/k\cdot v = 1/(k\cdot v + i\epsilon)$ is implied.   Inserting the Fourier transforms into the above expression then yields
\begin{eqnarray}	
\nonumber
\left.\mbox{Fig.}~\ref{fig:gluon1pt}(a)\right|_{{\cal O}(g^2)} &=&  g^2\sum_{\alpha,\beta\atop \alpha\neq\beta} \int_{\ell_\alpha,\ell_\beta} \mu_{\alpha,\beta}(k)\left[{c_\alpha\cdot c_\beta\over m_\alpha}{\ell_\alpha^2\over k\cdot v_\alpha}c_\alpha^a\left\{-v_\alpha\cdot v_\beta \left(\ell^\mu_\beta -{k\cdot\ell_\beta\over k\cdot v_\alpha} v^\mu_\alpha\right) +k\cdot v_\alpha v^\mu_\beta - k\cdot v_\beta v^\mu_\alpha\right\}  \right.\\
& & {} \left. \hspace{5.5cm}+ i(v_\alpha\cdot v_\beta)f^{abc} c^b_\alpha c^c_\beta  {\ell_\alpha^2\over k\cdot v_\alpha} v_\alpha^\mu \right].
\end{eqnarray}

The second contribution to ${\tilde J}^\mu_a(k)$ is from Fig.~\ref{fig:gluon1pt}(b) with static sources,
\be
\left. \mbox{Fig.}~\ref{fig:gluon1pt}(b)\right|_{{\cal O}(g^2)} =  g^2\sum_{\alpha,\beta\atop\alpha\neq\beta} \int_{\ell_\alpha,\ell_\beta}\mu_{\alpha,\beta}(k) if^{abc} c^b_\alpha c^c_\beta\left[2(k\cdot v_\beta) v^\mu_\alpha -  (v_\alpha\cdot v_\beta)\ell^\mu_\alpha\right].
\ee
Combining we find
\begin{eqnarray}
\label{eq:YMradLO}
\nonumber
\left. {\tilde J}^\mu_a(k)\right|_{{\cal O}(g^2)}  &=& g^2\sum_{\alpha,\beta\atop\alpha\neq\beta} \int_{\ell_\alpha,\ell_\beta}\mu_{\alpha,\beta}(k) \left[{c_\alpha\cdot c_\beta\over m_\alpha}{\ell_\alpha^2\over k\cdot v_\alpha}c_\alpha^a\left\{-v_\alpha\cdot v_\beta \left(\ell^\mu_\beta -{k\cdot\ell_\beta\over k\cdot v_\alpha} v^\mu_\alpha\right) +k\cdot v_\alpha v^\mu_\beta - k\cdot v_\beta v^\mu_\alpha\right\} \right.\\
& & {}\left. \hspace{2cm} +  if^{abc} c^b_\alpha c^c_\beta \left\{2(k\cdot v_\beta) v^\mu_\alpha - (v_\alpha\cdot v_\beta) \ell_\alpha^\mu +(v_\alpha\cdot v_\beta){\ell_\alpha^2\over k\cdot v_\alpha}  v^\mu_\alpha\right\}
\right]
\end{eqnarray}
A consistency check of this result is that it obeys the Ward identity $k_\mu {\tilde J}^\mu_a(k)=0$ even for off-shell $k^\mu$.   This requires a cancellation between diagrams, which is only possible after the leading order solution for the time dependent charge $c^a(\tau)$ is inserted into Fig.~\ref{fig:gluon1pt}(a).

\section{Classical solutions in dilaton gravity}
\label{sec:grav}

We will compare our results obtained above to classical solutions in $d$-dimensional gravity coupled to a scalar $\phi$  and to dynamical point sources.   At the two-derivative level, the action is $S=S_g+S_{pp}$, where the bulk theory is
\be
\label{eq:gaction}
S_g= -2 m_{Pl}^{d-2}\int d^d x \sqrt{g} \left[R -(d-2) g^{\mu\nu}\partial_\mu\phi\partial_\nu\phi\right],
\ee
and for a single particle
\be
\label{eq:pp}
S_{pp} = -m\int d\tau e^\phi,
\ee
($d\tau^2 = g_{\mu\nu} dx^\mu dx^\nu$).    We refer to $\phi$ as the dilaton because of the form of its couplings to the massive particle.    As in ref.~\cite{berngrant}, the presence of the scalar field, as well as its normalization and choice of interactions with the point particle, is motivated by the observation that the gauge theory Feynman rules that we use have no explicit dependence on the spacetime dimensionality.    This is in contrast to pure general relativity, where even the graviton propagator depends on $d$.    Thus the double copy of pure Yang-Mills coupled to charges must contain additional degrees of freedom beyond those of Einstein gravity in order to cancel the $d$-dependence.  Note that in the calculations below, we only need to keep scalar couplings up to quadratic order, $e^\phi=1+\phi + {1\over 2!} \phi^2+\cdots$.

The equations of motion for each particle are
\be
\label{eq:geom}
{dp^\mu\over d\tau} = -\left[\Gamma^\mu{}_{\rho\nu} {v}^\rho  + (v^\mu \partial_\nu- v_\nu \partial^\mu)\phi\right] p^\nu.
\ee
whose solution is $p^\mu(\tau) = W^\mu{}_\nu(\tau) p^\nu(-\infty)$, where the gravitational analog of the Wilson line\footnote{The version of the  gravitational Wilson line that appears here is the one given, e.g., in~\cite{sean,mod} (here corresponding to the affine connection for the conformally rescaled metric ${\tilde g}_{\mu\nu}=e^{2\phi} g_{\mu\nu}$).    In ref.~\cite{Brandhuber:2008tf} a different (non-covariant) definition of the gravitational Wilson loop was found to be related to perturbative amplitude calculations in ${\cal N}=8$ supergravity.}
\be
W^\mu{}_\nu(\tau) = P\exp\left[-\int^\tau dx^\rho\left\{\Gamma^\mu{}_{\rho\nu} + \left(\delta_\rho{}^\mu\partial_\nu-g_{\rho\nu}\partial^\mu\right)\phi\right\}\right].
\ee

To set up the perturbative expansion, we write the metric as $g_{\mu\nu} = \eta_{\mu\nu} + h_{\mu\nu}$.    The formal solution $\langle h_{\mu\nu}\rangle$ to the equations of motion can be written in deDonder gauge, $\partial_\nu h^{\mu\nu} = {1\over 2} \partial^\mu h^\sigma{}_\sigma$, as
\be
\label{eq:hvT}
\langle h_{\mu\nu}\rangle(x) = {1\over 2 m_{Pl}^{d-2}} \int_k {e^{-ik\cdot x}\over k^2} \left[{\tilde T}_{\mu\nu}(k) -{1\over d-2}\eta_{\mu\nu} {\tilde T}^\sigma{}_\sigma(k)\right],
\ee
where ${\tilde T}^{\mu\nu}(x)$ is a conserved, $\partial_\mu {\tilde T}^{\mu\nu}=0$, but coordinate dependent pseudo-tensor that includes the energy-momentum of the particle sources, the dilaton, and gravity.   Distinct but physically equivalent definitions of ${\tilde T}^{\mu\nu}(x)$ for pure gravity can be found in textbooks~\cite{Weinberg,LL}.   Our definition is the one used in~\cite{GnR}, with ${\tilde T}^{\mu\nu}(x)$ proportional to the coefficient of the graviton tadpole term in the background field gauge effective action $\Gamma[h,\phi]$ for this theory,
\be
\Gamma[h,\phi] = -{1\over 2}\int d^d x {\tilde T}^{\mu\nu}(x) h_{\mu\nu}(x)+\cdots.
\ee
The quantity ${\tilde T}^{\mu\nu}(k)$ with $k^2=0$ then determines the classical field measured by observers at spatial infinity.    For example, in four dimensions, the dimensionless strain at retarded time $t$ measured by gravitational wave detectors placed at $r\rightarrow\infty$ is obtained by dotting ${\tilde T}^{ij}$ into a normalized,  transverse-traceless polarization tensor $\epsilon_{ij}(k)$
\be
h_{\pm}(t,{\vec n}) =  {4G_N\over r} \int {d\omega\over 2\pi} e^{-i\omega t} \epsilon^*{}_{\pm}^{ij}(k) {\tilde T}_{ij}(k).
\ee
This yields the angular pattern of helicity $\pm 2$ gravitational waves along direction ${\vec n}={\vec k}/|{\vec k}|$ seen by observers far from the sources.   In any number of dimensions, the radiated energy-momentum is
\be
\Delta P^\mu = {1\over 4 m_{Pl}^{d-2}}\int_k (2\pi)\theta(k^0) \delta(k^2) k^\mu \left|\epsilon^*_{\rho\sigma}(k) {\tilde T}^{\rho\sigma}(k)\right|^2.
\ee

The full (non-perturbative) equation of motion for the scalar field is 
\be
\Box\phi(x) = -{1\over 4 m_{Pl}^{d-2}(d-2)}\sum_\alpha m_\alpha \int d\tau_\alpha e^{\phi(x)} {\delta(x-x_\alpha)\over\sqrt{g}},
\ee
whose solution we express as
\be
\label{eq:asdef}
\langle\phi\rangle(x) = -{1\over 2 m_{Pl}^{(d-2)\over 2}(d-2)^{1/2}} \int_k {e^{-i k\cdot x}\over k^2} {\cal A}_s(k).
\ee
For on-shell momentum $k^2=0$, ${\cal A}_s(k)$ can be interpreted as the (canonically normalized) semi-classical probability amplitude for scalar emission by the point sources.    From a strictly classical point of view, ${\cal A}_s(k)$ with on-shell $k^\mu$ determines the radiation field measured by detectors at future null infinity.

\subsection{The leading order radiation fields}

\begin{figure}
\centering
\includegraphics[scale=0.3]{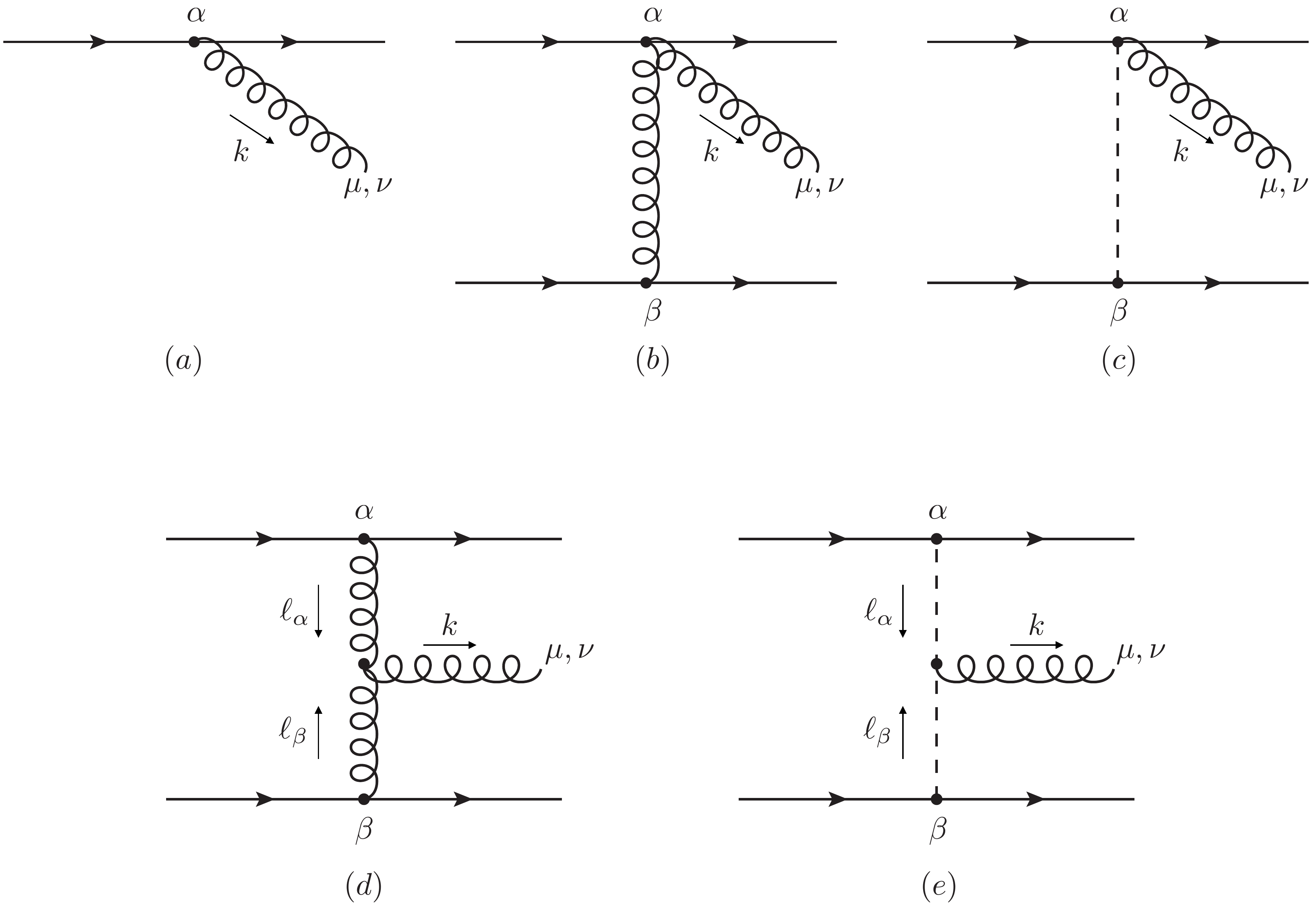}
\caption{\label{fig:graviton1pt}  Leading order Feynman diagrams in the perturbative expansion of ${\tilde T}^{\mu\nu}(k)$.}
\end{figure}

\begin{figure}
\centering
\includegraphics[scale=0.3]{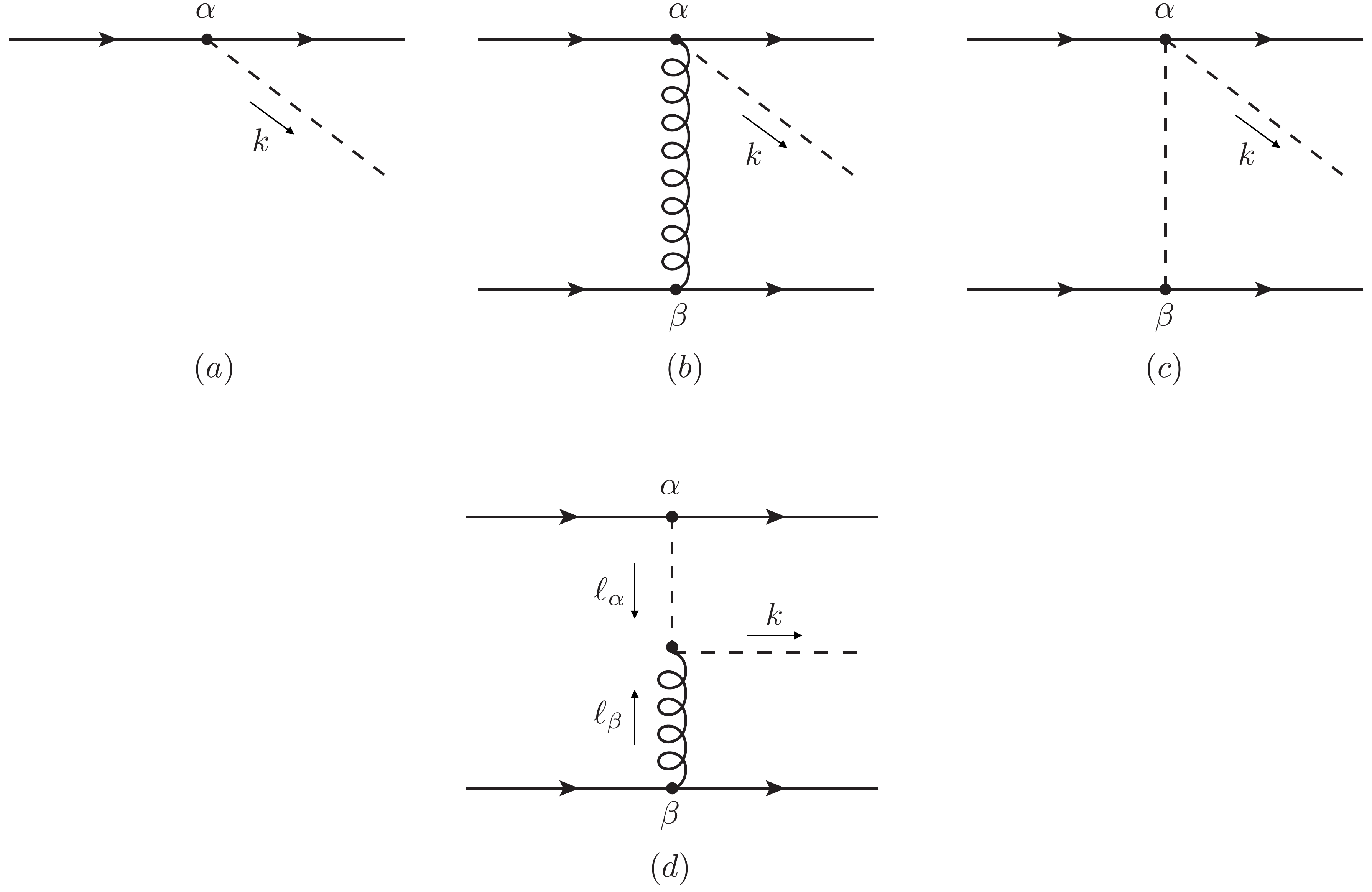}
\caption{\label{fig:scalar1pt}Leading order Feynman diagrams in the perturbative expansion of the scalar emission amplitude ${\cal A}_s(k)$.}
\end{figure}

We now consider a set of particles $\alpha=1,\ldots,N$ moving along trajectories $x_\alpha=b_\alpha + v_\alpha\tau + z_\alpha(\tau)$ subject to initial conditions $z_\alpha(\tau\rightarrow-\infty)=0$.    As in the Yang-Mills case, we focus on perturbative solutions, in which case ${\tilde T}^{\mu\nu}(k)$ and ${\cal A}_s(k)$ can be calculated in terms of the Feynman diagrams in Figs.~\ref{fig:graviton1pt},\ref{fig:scalar1pt}.  We follow the Feynman rule conventions given in~\cite{GnR}.   For example, the graviton propagator for internal lines is ${i\over m_{Pl}^{d-2}} P_{\mu\nu\rho\sigma}/k^2$, where the tensor structure is given by
\be
P_{\mu\nu\rho\sigma} = {1\over 2} \left[\eta_{\mu\rho}\eta_{\nu\sigma} + \eta_{\mu\sigma}\eta_{\nu\rho} -{2\over d-2} \eta_{\mu\nu}\eta_{\rho\sigma}\right].
\ee
For particles of comparable mass and energy $E\gtrsim m$, perturbation theory is valid in the kinematic regime
\be
\label{eq:geps}
\epsilon =  {\Gamma(d/2-3/2)\over (4\pi)^{(d-1)/2}}{E\over m_{Pl}^{d-2} b_{\alpha\beta}^{d-3}}\ll 1.
\ee   
Even though the Feynman diagrams in Fig.~\ref{fig:graviton1pt}, \ref{fig:scalar1pt} scale as definite powers of $1/m_{Pl}^{d-2},$ they do not exhibit manifest power counting in $\epsilon$.    Explicit $\epsilon$ scaling can be achieved by constructing an effective field theory for  the large impact parameter limit, but we do not attempt to do so here.   Again, we make no assumptions about the magnitude of the dimensionless quantity $\omega b$.

At leading order in perturbation theory, the particles travel undeflected, and source static fields that can be calculated from the diagrams in Fig.~\ref{fig:graviton1pt}(a), Fig.~\ref{fig:scalar1pt}(a) with $z^\mu_\alpha=0$.   The results are 
\be
\label{eq:LOmetric}
\langle h_{\mu\nu}\rangle(x) = {1\over 2 m_{Pl}^{d-2}}\sum_\alpha m_\alpha\int_\ell {e^{-i\ell\cdot(x-b_\alpha)}\over \ell^2}(2\pi)\delta(\ell\cdot v_\alpha)\left(v_{\alpha\mu} v_{\alpha\nu}  - {1\over d-2} \eta_{\mu\nu}\right),
\ee
\be
\langle\phi\rangle(x) = {1\over 4 m_{Pl}^{d-2}(d-2)}\sum_\alpha m_\alpha\int_\ell {e^{-i\ell\cdot (x-b_\alpha)}\over \ell^2}(2\pi)\delta(\ell\cdot v_\alpha).
\ee
Because the sources are static at lowest order, these solutions do not contain radiation.    However, they give rise to equations of motion
\be
\label{eq:gfma}
{d^2 z^\mu_\alpha\over d\tau^2} = {i\over 2 m^{d-2}_{Pl}}\sum_{\beta\neq\alpha} m_\beta \int_\ell (2\pi)\delta(\ell\cdot v_\beta) {e^{i\ell\cdot(b_{\alpha\beta}+v_\alpha\tau)}\over\ell^2}\left[{1\over 2}(v_\alpha\cdot v_\beta)^2 \ell^\mu -(\ell\cdot v_\alpha)\left((v_\alpha\cdot v_\beta) v^\mu_\beta - {v^\mu_\alpha\over 2(d-2)}\right)\right].
\ee
Thus the leading order Wilson line along the path of particle $\alpha$ is given by
\be
[\ln W(\tau\rightarrow\infty)]^\mu{}_\nu  = {i\over 4 m_{Pl}^{d-2}}\sum_{\beta\neq\alpha}m_\beta \int_{\ell_\alpha,\ell_\beta} \mu_{\alpha,\beta}(0) \ell^2_\alpha (v_\alpha\cdot v_\beta) \left(\ell_\alpha^\mu v_{\beta\nu} - \ell_{\alpha\nu} v^\mu_\beta\right).
\ee

Given these results, we can now obtain the leading order radiation amplitude ${\tilde T}^{\mu\nu}(k)$.   First plug in the solution $z^\mu_\alpha$ to Eq.~(\ref{eq:gfma}) into Fig.~\ref{fig:graviton1pt}(a),
\be
\mbox{Fig.}~\ref{fig:graviton1pt}(a) =\sum_\alpha e^{i k\cdot b_\alpha} m_\alpha\left(-ik\cdot v_\alpha\right)\left[ v_\alpha^\mu z_\alpha^\nu(\omega) + v_\alpha^\nu z^\mu_\alpha(\omega) - v_\alpha^\mu v_\alpha^\nu\left({k\cdot z_\alpha(\omega)\over k\cdot v_\alpha} + v_\alpha \cdot z_\alpha(\omega)\right)\right]_{\omega=k\cdot v_\alpha},
\ee
to obtain
\begin{eqnarray}
\nonumber
\mbox{Fig.}~\ref{fig:graviton1pt}(a) &=& {1\over 2 m_{Pl}^{d-2}} \sum_{\alpha,\beta\atop\alpha \neq\beta} m_\alpha m_\beta \int_{\ell_\alpha,\ell_\beta} \mu_{\alpha,\beta}(k) {\ell^2_\alpha\over k\cdot v_\alpha}\left[{1\over 2} (v_\alpha\cdot v_\beta)^2 \left(v^\mu_\alpha\ell^\nu_\beta + v^\nu_\alpha\ell^\mu_\beta -{k\cdot\ell_\beta\over k\cdot v_\alpha} v^\mu_\alpha v^\nu_\alpha\right) \right.\\
\nonumber\\
& & {} \left.\hspace{2.5cm}+ (v_\alpha\cdot v_\beta)\left({1\over 2} (v_\alpha\cdot v_\beta) k\cdot v_\alpha + k\cdot v_\beta\right) v_\alpha^\mu v_\alpha^\nu - (v_\alpha\cdot v_\beta) (k\cdot v_\alpha)(v^\mu_\alpha v^\nu_\beta + v^\nu_\alpha v^\mu_\beta)\right].
\end{eqnarray}
Note that we have dropped terms proportional to $\delta(k\cdot v_\alpha)$ which, either for massive or massless particles, do not contribute to the radiation field\footnote{In the massless case, the argument of the delta function can be non-zero if $k^\mu$ points along the direction of a particle momentum.    However, because the $\delta(k\cdot v_\alpha)$ term is proportional to $p_\alpha^\mu p_\alpha^\nu$, dotting the amplitude into an external graviton, or taking a trace to project onto scalar radiation gives a vanishing result.} .   At this order in perturbation theory, we must also compute the diagrams in Fig.~\ref{fig:graviton1pt}(b)-(e) at zero deflection, whose respective contributions to ${\tilde T}^{\mu\nu}(k)$ are
\begin{eqnarray}
\mbox{Fig.}~\ref{fig:graviton1pt}(b) &=&-{1\over 4 m^{d-2}_{Pl}}\sum_{\alpha,\beta\atop\alpha\neq\beta}m_\alpha m_\beta \int_{\ell_\alpha,\ell_\beta} \mu_{\alpha,\beta}(k)\left[(v_\alpha\cdot v_\beta)^2 - {1\over d-2}\right] \ell_\alpha^2 v^\mu_\alpha v^\nu_\alpha,\\
\mbox{Fig.}~\ref{fig:graviton1pt}(c) &=& {1\over 4 m^{d-2}_{Pl}}\sum_{\alpha,\beta\atop\alpha\neq\beta} m_\alpha m_\beta \int_{\ell_\alpha,\ell_\beta} \mu_{\alpha,\beta}(k) {\ell_\alpha^2 v^\mu_\alpha v^\nu_\alpha\over d-2},
\end{eqnarray}
and
\begin{eqnarray}
\nonumber
\mbox{Fig.}~\ref{fig:graviton1pt}(d) &=& {1\over 4 m^{d-2}_{Pl}}\sum_{\alpha,\beta\atop \alpha\neq\beta} m_\alpha m_\beta \int_{\ell_\alpha,\ell_\beta} \mu_{\alpha,\beta}(k)\left[\left((v_\alpha\cdot v_\beta)^2 - {1\over d-2}\right)\left(2\ell^\mu_\alpha \ell^\nu_\alpha + \ell^\mu_\alpha \ell^\nu_\beta\right)\right.\\
\nonumber\\
\nonumber
& & \hspace{1cm} + \eta^{\mu\nu}\left\{v_\alpha\cdot v_\beta (k\cdot v_\alpha) (k\cdot v_\beta) -{2 (k\cdot v_\alpha)^2\over d-2} -{1\over 2}\left((v_\alpha\cdot v_\beta)^2 - {1\over d-2}\right)\ell_\alpha^2\right\}\\
\nonumber\\
\nonumber
& & \hspace{1cm}+ 2 \left({k\cdot v_\alpha\over d-2} - (v_\alpha\cdot v_\beta) k\cdot v_\beta\right) \left(v_\alpha^\mu\ell^\nu_\alpha + v_\alpha^\nu \ell^\mu_\alpha\right) + {2 k\cdot v_\beta\over d-2} \left(\ell^\mu_\alpha  v^\nu_\beta + \ell^\nu_\alpha  v^\mu_\beta\right)\\
\nonumber\\
& & \left. \hspace{1cm}+ 2\left((k\cdot v_\beta)^2 - {\ell_\alpha^2\over d-2}\right) v^\mu_\alpha v^\nu_\alpha + \left\{ \ell_\alpha^2 v_\alpha\cdot v_\beta - k\cdot v_\alpha k\cdot v_\beta\right\}\left(v^\mu_\alpha v^\nu_\beta + v^\nu_\alpha v^\mu_\beta\right)\right].
\end{eqnarray}
To get the result in Fig.~\ref{fig:graviton1pt}(d), we have used the background field gauge three-graviton interaction vertex, whose explicit form can be found, e.g., in~\cite{GnR}.    The remaining contribution to ${\tilde T}^{\mu\nu}(k)$ at this order in the interactions is
\be
\mbox{Fig.}~\ref{fig:graviton1pt}(e) = -{1\over 4 m^{d-2}_{Pl}(d-2)}\sum_{\alpha,\beta\atop\alpha\neq\beta} m_\alpha m_\beta \int_{\ell_\alpha,\ell_\beta} \mu_{\alpha,\beta}(k) \left[\ell^\mu_\alpha \ell^\nu_\beta - {1\over 2} \eta^{\mu\nu} \ell_\alpha\cdot\ell_\beta  \right].
\ee
In all these equations, we have dropped terms that vanish when $k^\mu$ is on-shell, as these do not contribute to the asymptotic field at $r\rightarrow\infty$.   However, we have checked that the sum of the diagrams Fig.~\ref{fig:graviton1pt}(a)-(e) obeys the Ward identity $k_\mu {\tilde T}^{\mu\nu}(k)=0$ even for $k^\mu$ off-shell.    In order to compare to the analogous Yang-Mills results, we will only focus on the components of ${\tilde T}^{\mu\nu}(k)$ which contribute to the radiation field at infinity.    In particular, the canonically normalized graviton emission amplitude simplifies to
\begin{eqnarray}
\label{eq:gos}
\nonumber
{\cal A}_g(k) &=& -{1\over 2 m_{Pl}^{{(d-2)}/2}} \epsilon^*_{\mu\nu}(k){\tilde T}^{\mu\nu}(k)= -{\epsilon^*_{\mu\nu}(k)\over 8 m^{3(d-2)/2}_{Pl}}\sum_{\alpha,\beta\atop\alpha\neq\beta} m_\alpha m_\beta \int_{\ell_\alpha,\ell_\beta} \mu_{\alpha,\beta}(k)\left[(v_\alpha\cdot v_\beta)^2 \ell_\alpha^\mu \ell^\nu_\alpha\right.\\
\nonumber
\\
\nonumber 
& & \hspace{2.5cm}+ (v_\alpha\cdot v_\beta)\eta^{\mu\nu}\left\{{1\over 2} (v_\alpha\cdot v_\beta) \ell_\alpha^2 + (k\cdot v_\alpha) (k\cdot v_\beta)\right\}\\
\nonumber
\\
\nonumber
& & \hspace{2.5cm}- 2 (v_\alpha\cdot v_\beta)\left((v_\alpha\cdot v_\beta){\ell_\alpha^2\over k\cdot v_\alpha} + 2 k\cdot v_\beta\right) \ell^\mu_\alpha v_\alpha^\nu
 -2\left((k\cdot v_\alpha) (k\cdot v_\beta) +  (v_\alpha\cdot v_\beta) \ell_\alpha^2 \right) v_\alpha^\mu v^\nu_\beta\\
\nonumber
\\
& &\left.  \hspace{2.5cm} + \left\{(v_\alpha\cdot v_\beta){\ell_\alpha^2\over (k\cdot v_\alpha)^2}\left((v_\alpha\cdot v_\beta) k\cdot\ell_\alpha + 2 (k\cdot v_\alpha) (k\cdot v_\beta)\right) + 2 (k\cdot v_\beta)^2\right\}  v_\alpha^\mu v_\alpha^\nu\right],
\end{eqnarray}
where we have only assumed that the polarization tensor obeys the deDonder gauge condition $k^\mu \epsilon_{\mu\nu}(k) = {1\over 2} k_\nu \epsilon^\sigma{}_\sigma(k)$.   Note in particular that, by construction, all explicit dependence on the spacetime dimensionality cancels in this on-shell quantity.   This would not be true for the non-radiative components of the solution at this order, and it would not be true of the radiation amplitude in pure gravity (diagrams (a), (b), (d) in Fig.~\ref{fig:graviton1pt}).   This cancellation is what dictates the choice of scalar interactions, and is going to be important later when we discuss double copy relations between the Yang-Mills solution and the result in Eq.~(\ref{eq:gos}).

We can use the same methods to calculate the amplitude for dilaton emission from the classical system.   Inserting the solution to Eq.~(\ref{eq:gfma}) into Fig~\ref{fig:scalar1pt}(a), we obtain the following contribution to ${\cal A}_s(k)$ (defined in Eq.~(\ref{eq:asdef})),
\begin{eqnarray}
\nonumber
\mbox{Fig.}~\ref{fig:scalar1pt}(a) &=& -{1\over 4 m_{Pl}^{3(d-2)/2} (d-2)^{1/2}}\sum_{\alpha,\beta\atop\alpha\neq\beta} m_\alpha m_\beta \int_{\ell_\alpha,\ell_\beta} \mu_{\alpha,\beta}(k)  {\ell^2_\alpha\over (k\cdot v_\alpha)^2}   (v_\alpha\cdot v_\beta)\left[(k\cdot v_\alpha) (k\cdot v_\beta)\right.\\
\nonumber\\
& & \left. {} \hspace{6.5cm}-{1\over 2} (v_\alpha\cdot v_\beta) \left(k\cdot \ell_\beta + (k\cdot v_\alpha)^2\right)\right],
\end{eqnarray}
and working at zero deflection,
\begin{eqnarray}
\mbox{Fig.}~\ref{fig:scalar1pt}(b)  &=& -{1\over 8 m_{Pl}^{3(d-2)/2} (d-2)^{1/2}}\sum_{\alpha,\beta\atop\alpha\neq\beta} m_\alpha m_\beta  \int_{\ell_\alpha,\ell_\beta} \mu_{\alpha,\beta}(k) \left[(v_\alpha\cdot v_\beta)^2 - {1\over d-2}\right] \ell^2_\alpha,\\
\nonumber\\
\mbox{Fig.}~\ref{fig:scalar1pt}(c)  &=& -{1\over 8 m_{Pl}^{3(d-2)/2} (d-2)^{1/2}}\sum_{\alpha,\beta\atop\alpha\neq\beta} m_\alpha m_\beta \int_{\ell_\alpha,\ell_\beta} \mu_{\alpha,\beta}(k)\, {\ell^2_\alpha\over d-2},\\
\nonumber\\
\mbox{Fig.}~\ref{fig:scalar1pt}(d)  &=& -{1\over 4 m_{Pl}^{3(d-2)/2} (d-2)^{1/2}}\sum_{\alpha,\beta\atop\alpha\neq\beta} m_\alpha m_\beta \int_{\ell_\alpha,\ell_\beta} \mu_{\alpha,\beta}(k) \, (k\cdot v_\alpha)^2. 
\end{eqnarray}
In particular, the diagram Fig.~\ref{fig:scalar1pt}(c) probes the coefficient of the quadratic coupling of the scalar to the particles written in Eq.~(\ref{eq:pp}).  The sum of these diagrams then gives the scalar radiation amplitude
\begin{eqnarray}
\nonumber
\label{eq:scalaramp}
{\cal A}_s(k) &=& -{1\over 8 m_{Pl}^{3(d-2)/2} (d-2)^{1/2}}\sum_{\alpha,\beta\atop\alpha\neq\beta} m_\alpha m_\beta \int_{\ell_\alpha,\ell_\beta} \mu_{\alpha,\beta}(k)\left[{(v_\alpha\cdot v_\beta)\ell^2_\alpha\over (k\cdot v_\alpha)^2}\left\{(v_\alpha\cdot v_\beta) k\cdot\ell_\alpha + 2 (k\cdot v_\alpha) (k\cdot v_\beta)\right\} + 2 (k\cdot v_\alpha)^2  \right].\\
\end{eqnarray}

\section{Color-kinematics correspondence}
\label{sec:doublecopy}

We now show that there is a color-kinematics relation between the perturbative Yang-Mills observables in sec.~\ref{sec:YM} and the corresponding ones in the  scalar-gravity theory constructed in sec.~\ref{sec:grav}.    The possibility of such a correspondence is well motivated by work on scattering amplitudes, going back to the KLT relations and more recently the BCJ double copy of gauge theory.   

As in the BCJ case, the connection between gauge and gravity observables consists of making certain color-to-kinematics substitutions.   To recover gravity from our Yang-Mills results, we first replace the initial color charge of each particle with a second copy of its initial momentum
\be
\label{eq:sub1}
c^a \rightarrow p^\mu.
\ee
This substitution is motivated by the structural similarity between the respective classical equations of motion:   in gauge theory, charge $c^a(\tau)$ is parallel transported in color space, while in gravity $p^\mu(\tau)$ also obeys a parallel transport equation, Eq.~(\ref{eq:geom}), generated by the affine connection associated with the Weyl re-scaled metric  ${\tilde g}_{\mu\nu} = e^{2 \phi} g_{\mu\nu}$.     The replacement Eq.~(\ref{eq:sub1}) is also similar to the identification $c^a\rightarrow m$ in the Kerr-Schild double copy proposal of~\cite{Monteiro:2014cda}.     Note that under the replacement in Eq.~(\ref{eq:sub1}), the trivial (order $\epsilon^0$) gluon field in Eq.~(\ref{eq:LOA}) \emph{does not} map onto the gravitational solution in Eq.~(\ref{eq:LOmetric}), except in the case of highly boosted sources.    The fact that the massless limit double copies in this way is consistent with observations made in ref.~\cite{Saotome:2012vy}, which constructed these solutions indirectly, by resumming the eikonal limit of QCD and using the BCJ relations to make contact with classical gravity.    The special case of massless sources and their classical double copy is discussed in more detail below in sec.~\ref{sec:massless}.

Even though the (gauge dependent) leading order solutions in Eq.~(\ref{eq:LOA}) and Eq.~(\ref{eq:LOmetric}) are only related for $m=0$ particles, we find that gauge-invariant classical observables, in particular the transverse radiation field at $r\rightarrow\infty$, does obey a double copy relation even in the more general case of massive sources.    In order to see this relation, we have to compare the solutions at the next order in perturbation theory, where radiation first shows up.    At this order in the expansion, it becomes necessary to introduce a substitution rule for the color structure $f^{abc}$ on the gauge theory side.    In our gauge theory results, every term containing a factor of $f^{abc}$ can be associated with an expression involving incoming momenta $q_{1,2,3}$.   As in BCJ, it is natural to replace the structure constants with a second copy of the kinematic factor appearing in the 3-gluon Feynman vertex,
\be
\label{eq:sub2}
i f^{a_1 a_2 a_3}\rightarrow \Gamma^{\nu_1\nu_2\nu_3}(q_1,q_2,q_3) =-{1\over 2} \left[\eta^{\nu_1\nu_3}(q_1-q_3)^{\nu_2} +\eta^{\nu_1\nu_2} (q_2-q_1)^{\nu_3} + \eta^{\nu_2\nu_3}(q_3-q_2)^{\nu_1}\right],
\ee
with $q_1+q_2+q_3=0$.
Finally, to compare to gravity, we introduce the replacement rule
\be
\label{eq:sub3}
g\rightarrow {1\over 2 m_{Pl}^{d/2-1}}.
\ee

Given these rules, we can now determine the gravitational double copy of the emission amplitude ${\cal A}(k) = \epsilon^a_\mu(k) \left[g {\tilde J}^a_{\mu}(k)\right]$.   We replace the gluon polarization vector,
\be
\epsilon^a_\mu(k)\rightarrow \epsilon_\mu(k) {\tilde\epsilon}_\nu(k),
\ee
with independent photon polarizations $\epsilon_\mu(k)$,  ${\tilde \epsilon}_\mu(k)$.   Making these substitutions, the double copy ${\hat{\cal A}}_{\mu\nu}(k)$ of the Yang-Mills amplitude is given by
\be
\epsilon^a_\mu(k) \left[g {\tilde J}^a_{\mu}(k)\right] \rightarrow \epsilon_\mu(k) {\tilde\epsilon}_\nu(k) {\hat{\cal A}}_{\mu\nu}(k),
\ee
which is only well-defined up to terms that vanish when dotted into the external on-shell polarizations.     Using this gauge freedom, the double copy amplitude with $k^2=0$ can be taken to be, from Eq.~(\ref{eq:YMradLO}),
\begin{eqnarray}
\nonumber
{\hat{\cal A}}^{\mu\nu}(k) &=& - \sum_{{\alpha,\beta}\atop{{\alpha\neq\beta}}} {m_\alpha m_\beta\over 8 m_{Pl}^{3(d-2)/2}} \int_{\ell_\alpha,\ell_\beta} \mu_{\alpha,\beta}(k)\left[ {({v_\alpha\cdot v_\beta})\ell_\alpha^2\over k\cdot v_\alpha} v_\alpha^\nu\left\{(v_\alpha\cdot v_\beta)\left({1\over 2}(\ell_\beta-\ell_\alpha)^\mu - {k\cdot\ell_\beta\over k\cdot v_\alpha} v_\alpha^\mu\right)+(k\cdot v_\beta) v^\mu_\alpha-(k\cdot v_\alpha) v^\mu_\beta\right\}\right.\\
\nonumber\\
\nonumber
& &\left. + {1\over 2} \left\{2 (k\cdot v_\beta) v_\alpha^\nu - 2 (k\cdot v_\alpha) v_\beta^\nu+(v_\alpha\cdot v_\beta) (\ell_\beta-\ell_\alpha)^\nu\right\}\left\{2 (k\cdot v_\beta) v_\alpha^\mu - (v_\alpha\cdot v_\beta) \ell_\alpha^\mu+{(v_\alpha\cdot v_\beta)\ell_\alpha^2\over k\cdot v_\alpha} v^\mu_\alpha\right\}\right].\\
\end{eqnarray}
To obtain, this, we have added pure gauge terms proportional to $k^\nu$ whose specific form has been chosen such that, for on-shell $k^2=0$, the Ward identity is obeyed in each Lorentz index,
\be
k^\mu {\hat{\cal A}}_{\mu\nu}(k)= k^\nu {\hat{\cal A}}_{\mu\nu}(k)=0.
\ee

The double copy amplitude ${\hat{\cal A}}_{\mu\nu}(k)$ simultaneously encodes radiation in the graviton and scalar channels.    To extract each process, we decompose the product $\epsilon^\mu {\tilde\epsilon}^\nu$ into irreducible $SO(d-2)$ massless little group representations.   In unitary gauge, with $\epsilon_0(k)={\tilde \epsilon}_0(k)=0,$ the spatial components decompose as 
\be
\epsilon_i(k) {\tilde\epsilon}_j(k)=\epsilon_{ij}(k) + a_{ij}(k) - {{\epsilon}(k)\cdot{{\tilde\epsilon}}(k)\over d-2} h_{ij}(k),
\ee
where the scalar is proportional to $h_{ij}(k)=\delta_{ij} - {k_i k_j}/{\vec k}^2$ and the transverse traceless graviton is 
\be
\label{eq:gpol}
\epsilon_{ij}(k) = {1\over 2} \epsilon_i(k){\tilde\epsilon}_j(k) + {1\over 2} \epsilon_j(k){\tilde\epsilon}_i(k) + {{\epsilon}(k)\cdot{{\tilde\epsilon}}(k)\over d-2} h_{ij}(k).
\ee
In principle, there is also the anti-symmetric mode $a_{ij}(k) = \left(\epsilon_i(k){\tilde\epsilon}_j(k) - \epsilon_j(k){\tilde\epsilon}_i(k)\right)/2$, associated with the existence of a two-form gauge field $B_{\mu\nu}$ in the Yang-Mills double copy.   However, by direct calculation it is easy to see that $a_{ij}(k) {\hat{\cal A}}^{ij}(k)=0$, so that this field is not radiated by the point sources.     We will comment further on the role played by the $B_{\mu\nu}$ field in sec.~\ref{sec:conc}.   The double copy ${\hat{\cal A}}_{\mu\nu}(k)$ contains a canonically normalized scalar emission amplitude  of the form    
\be
-{h^{ij}(k) {\hat{\cal A}}_{ij}(k)\over\sqrt{h_{mn} h^{mn}(k)}} = {\eta^{\mu\nu} {\hat{\cal A}}_{\mu\nu}(k)\over (d-2)^{1/2}}= {\cal A}_s(k),
\ee
which reproduces the result Eq.~(\ref{eq:scalaramp}) obtained in the gravity theory.    In the graviton channel, the double copy amplitude is
\begin{eqnarray}
\epsilon_{ij}(k) {\hat{\cal A}}^{ij}(k) 
&=& {1\over 2} \epsilon_\mu(k){\tilde\epsilon}_\nu(k)\left[{\hat{\cal A}}^{\mu\nu}(k)+{\hat{\cal A}}^{\nu\mu}(k)\right]-{{\epsilon}(k)\cdot{{\tilde\epsilon}}(k)\over d-2}{\hat{\cal A}}^\mu{}_\mu(k).
\end{eqnarray}
For graviton polarization of the form given in Eq.~(\ref{eq:gpol}), this expression agrees with the asymptotic radiation field in Eq.~(\ref{eq:gos}),
\be
\epsilon_{ij}(k) {\hat{\cal A}}^{ij}(k)  =  -{1\over 2 m_{Pl}^{(d-2)/2}}\epsilon_{\mu\nu}(k) {\tilde T}^{\mu\nu}(k),
\ee 
computed directly in perturbative gravity.

\subsection{The double copy of a static point source}
\label{sec:ptgmunu}

Taken at face value, our results in the previous section imply that the double copy of an isolated color charge is an object that sources both the graviton and dilaton.     We can use the results in sec.~\ref{sec:grav} to calculate the long distance gravitational field of this object.   This same method was first used by Duff~\cite{Duff} to calculate the asymptotic four-dimensional Schwarzschild metric to second order in powers of $r_s/r\ll 1$ and in~\cite{GnR} to order $(r_s/r)^3$, where $r_s$ is the Schwarzschild radius of the source.  

For a single particle with $b^\mu=0$, the leading order field as an expansion in powers of $(r_s/r)^{d-3}\ll 1$ can be read off  directly from Eq.~(\ref{eq:LOmetric}).  In a deDonder coordinate system, with $\eta^{\mu\nu}\Gamma^\lambda_{\mu\nu}=0$,
\be
\label{eq:h1}
h_{\mu\nu}(x)= {1\over d-3}\left({r_s\over r}\right)^{d-3} \left(\eta_{\mu\nu}-(d-2) v_\mu v_\nu\right),
\ee
with $v^\mu$ the source's velocity.   In terms of the spatial coordinates $x^\mu_\perp = x^\mu - (v\cdot x) v^\mu$, the radial variable is defined to be $r = \sqrt{-x_\perp^2}$.    At this order in perturbation theory, the solution coincides with the $d$-dimensional Schwarzschild metric, and by defining a new radial coordinate 
\be
\rho =  r\left[1+{1\over 2 (d-3)}\left({r_s\over r}\right)^{d-3}+\cdots\right],
\ee
the gravitational field of an isolated particle can be put in the standard Schwarzschild form
\be
\label{eq:std}
ds^2 = f(\rho) dt^2  - g(\rho) d\rho^2 - \rho^2 d\Omega_{d-2}^2,
\ee
\begin{eqnarray}
f(\rho)   &=&  1-\left({r_s\over \rho}\right)^{d-3}+\cdots,\\
 g(\rho) &=& 1+\left({r_s\over \rho}\right)^{d-3}+\cdots. 
\end{eqnarray}

At the next order in the expansion, the backreaction of the scalar field profile on the metric must be taken into account.    Summing up the energy-momentum of the graviton and $\phi$, we find at second order in perturbation theory\footnote{This result can be obtained from the diagrams of Fig.~\ref{fig:graviton1pt}(d), Fig.~\ref{fig:graviton1pt}(e) in the case of a single particle.}
\begin{eqnarray}
\nonumber
{\tilde T}^{\mu\nu}(k) &=& {m^2\over 8 (d-2) m_{Pl}^{d-2}} (2\pi)\delta(k\cdot v) \int_\ell (2\pi)\delta(\ell\cdot v) {1\over \ell^2}{1\over (\ell+k)^2} \left[2(d-2)\ell^\mu \ell^\nu + (d-2) (\ell^\mu k^\nu +\ell^\nu k^\mu) \right.\\
\nonumber\\
& & \left.+2 (d-3) k^\mu k^\nu  -{1\over 2}(3d-10) k^2 \eta^{\mu\nu}+ 2 (d-4) k^2 v^\mu v^\nu\right].
\end{eqnarray}
The correction to the metric can now be extracted from Eq.~(\ref{eq:hvT}).    In order to calculate $h_{\mu\nu}(x)$ we first do the integral over momentum $\ell$ by standard one-loop Feynman parameter methods, and then the Fourier transform over $k$ using the identity
\be
\int_k (2\pi)\delta(k\cdot v) {{e^{-i k\cdot  x} \over (-{k}^2)^\alpha}}= {1\over (4\pi)^{{d-1\over 2}}} {\Gamma(d/2-\alpha-1/2)\over \Gamma(\alpha)}\left({2\over r}\right)^{d-2\alpha-1}.
\ee
The result is that $h^{(2)}_{\mu\nu}(x)$, the order $(r_s/r)^{2(d-3)}$ correction to the metric, is given by 
\be
h^{(2)}_{\mu\nu}(x) = {1\over 4} {d-2\over (d-3)^2} \left({r_s\over r}\right)^{2(d-3)}\left[-{2 d^2-13d+24\over (d-2) (d-5)} \eta_{\mu\nu} + {2 d^2 -16 d +33\over d-5} v_\mu v_\nu -  {3(d-3)^2\over d-5} {x^\perp_\mu x^\perp_\nu\over r^2}\right].
\ee
By re-defining the radial coordinate, 
\be
\rho^2 = r^2\left[1+{1\over (d-3)}\left({r_s\over r}\right)^{d-3}-{2d^2-13d+24\over 4(d-3)^2 (d-5)}\left({r_s\over r}\right)^{2(d-3)} +\cdots\right],
\ee
the metric including second order corrections can again be put in the form given in Eq.~(\ref{eq:std}), where now
\begin{eqnarray}
\label{eq:2f}
f(\rho)   &=&  1-\left({r_s\over \rho}\right)^{d-3}+0\cdot \left({r_s\over \rho}\right)^{2(d-3)} +\cdots, \\
\label{eq:2g}
 g(\rho) &=& 1+\left({r_s\over \rho}\right)^{d-3}+ \left(1-{1\over 4(d-3)}\right)\left({r_s\over \rho}\right)^{2(d-3)} +\cdots. 
\end{eqnarray}
The second order corrections to $g(\rho)$ includes a pure gravity contribution that does not depend on $d$, and a scalar contribution with logarithmic UV divergences in $d=3$ dimensions\footnote{Such divergences renormalize the coefficients of non-minimal interactions between the particle worldline and $g_{\mu\nu}$, $\phi$~\cite{GnR}.}.   Note that in the large $d$ limit~\cite{Strominger:1981jg,BjerrumBohr:2003zd,Emparan:2013moa}, the scalar decouples and we recover the $d$-dimensional Schwarzschild solution to second order in perturbation theory.    It is also straightforward to calculate the scalar field profile.  In deDonder coordinates, the result is
\be
\phi(r) = {1\over 2 (d-3)} \left({r_s\over r}\right)^{d-3} + 0\cdot \left({r_s\over r}\right)^{2(d-3)}+\cdots,
\ee
while in Schwarzschild coordinates
\be
\phi(\rho) =  {1\over 2 (d-3)} \left({r_s\over \rho}\right)^{d-3} + {1\over 4 (d-3)}\left({r_s\over \rho}\right)^{2(d-3)} +\cdots.
\ee

There is a natural connection~\cite{Monteiro:2014cda} between certain solutions of the classical Yang-Mills equations and Kerr-Schild solutions to the Einstein equations in pure gravity.   According to this correspondence, the exact non-Abelian Coulomb field of a point color charge corresponds to the Schwarzschild solution of vacuum general relativity.      The results in Eq.~(\ref{eq:2g}) instead indicate that, because of  scalar charge, the double copy of the Coulomb field is neither a vacuum spacetime nor is it equivalent to Kerr-Schild by coordinate transformations.    To see this latter assertion, note that the general  $d$-dimensional spherically symmetric Kerr-Schild metric, $g_{\mu\nu}=\eta_{\mu\nu} +\chi(\rho) k_\mu k_\nu$, $k^\mu = (1,x^i/\rho)$, $\rho=|{\vec x}|$ can be put~\cite{MarkandAlec} in Schwarzschild form
\be
ds^2 = \left(1+\chi(\rho)\right)dt^2 - {d\rho^2 \over1+\chi(\rho)} -\rho^2 d\Omega_{d-2}^2
\ee
by a suitable choice of constant time slices.    By comparison with Eqs.~(\ref{eq:2f}),~(\ref{eq:2g}) we see that the perturbative double copy is not of Kerr-Schild form starting at second order in perturbation theory.

We are forced by this analysis to conclude that in the one-body sector, Kerr-Schild duality is at odds with the strictly perturbative approach in this paper,  except possibly in two physically interesting limits.    First, in the case of infinite spacetime dimensions~\cite{Strominger:1981jg,BjerrumBohr:2003zd,Emparan:2013moa}, it is clear from Eq.~(\ref{eq:gaction}) that the scalar field consistently decouples to all orders in perturbation theory.   In this case, the perturbative double copy of Yang-Mills is the $d\rightarrow\infty$ limit of pure general relativity, and our result in Eqs.~(\ref{eq:2f}),~(\ref{eq:2g}) reduces to the Schwarzschild vacuum.    The second limit is that of massless charges, which can be obtained from the results of this section by taking the ultrarelativistic limit $v^0=\gamma\rightarrow\infty$.    In that limit the linearized result in Eq.~(\ref{eq:h1}) becomes exact to all orders in powers of $(r_s/r)^{d-3}$ regardless of the dilaton couplings, and the gravitational field is the Aichelburg-Sexl shockwave~\cite{Aichelburg:1970dh}, which is one of the Kerr-Schild solutions considered in~\cite{Monteiro:2014cda}.

\subsection{Pure gravity and the double copy for massless particles}
\label{sec:massless}

In fact, for strictly massless particle sources, the perturbative classical double copy of Yang-Mills theory is pure Einstein gravity.    This is seen most transparently by introducing into the point particle Lagrangian an auxiliary ``einbein" $\eta(\lambda)$ whose role is to enforce invariance under reparametrizations $\lambda\rightarrow {\tilde\lambda}(\lambda)$ of the worldline (see e.g.~\cite{polchinski}).    In terms of this new degree of freedom, the action takes the form
\be
S_{pp} = -{1\over 2} \int d\lambda \left[{\eta^{-1}} e^{2 \phi} g_{\mu\nu} {d x^\mu\over d\lambda} {d x^\nu\over d\lambda}  + \eta m^2\right].
\ee
Varying the action with respect to $\eta$ then yields the constraint 
\be
\label{eq:constraint}
 e^{2 \phi} g_{\mu\nu} {d x^\mu\over d\lambda} {d x^\nu\over d\lambda} = m^2\eta^2,
\ee
which when inserted back into the Lagrangian reproduces the action $-m\int d\tau e^\phi$ assumed in sec.~\ref{sec:grav}.    In the einbein formulation, $\phi$ obeys the equation of motion
\be
\Box\phi = -{1\over 8 m_{Pl}^{d-2}(d-2)}\sum_\alpha  \int d\lambda \eta^{-1} e^{2\phi(x)}  g_{\mu\nu} {d x^\mu\over d\lambda} {d x^\nu\over d\lambda} {\delta(x-x_\alpha)\over\sqrt{g}}.
\ee

It is now straightforward to take $m=0$, in which case $x^\mu(\lambda)$ obeys the null geodesic equation in the metric $e^{2\phi} g_{\mu\nu}$, while  the constraint Eq.~(\ref{eq:constraint}) reduces the dilaton equation to $\Box \phi=0$.     Because null geodesics are preserved under Weyl re-scalings, the dilaton has no effect on the motion of the particles, and in any case the trivial solution with constant $\phi$ satisfies the equations of motion to all orders in perturbation theory.  Thus, if all the particles are massless, the scalar mode decouples and the double copy of the gauge theory configuration is automatically a solution of pure general relativity.

\section{Discussion and Outlook}
\label{sec:conc}

In this paper we have constructed perturbative classical solutions of Yang-Mills theory coupled to point charges and analogous solutions in a theory of a graviton and a massless scalar field.   Our results hold in any number of spacetime dimensions, and for any number of point sources.    By applying a simple set of BCJ-motivated replacement rules, encapsulated in Eqs.~(\ref{eq:sub1}),~(\ref{eq:sub2}),~(\ref{eq:sub3}), we are able to reproduce the gravitational and scalar fields detected by far away observers from the corresponding asymptotic gluon field in classical gauge theory.    We have focused on solutions corresponding to classical unbound trajectories, but our results can also be easily applied to obtain solutions with particles in non-relativistic bound orbits.

Several remarks are in order.      First, given their close relation to on-shell observables, it is perhaps not completely unexpected that the asymptotic classical gravity solutions we construct are related to gauge theory by rules similar to the BCJ transformations that hold in the case of the $S$-matrix.   However,  the double copy structure of classical solutions seems to differ in one important way from the relations for scattering amplitudes.    Namely, for scattering amplitudes, it is essential that matrix elements be put into explicit BCJ form, as a sum over groupings of Feynman diagrams whose numerators are related by the Jacobi identity of the Lie algebra.   In our case, the classical Yang-Mills radiation field is only linear in the structure constants, and aside from total anti-symmetry under index permutations,  algebraic relations obeyed by $f^{abc}$ seems to not play any role.    Going to the next order in perturbation theory, the classical solutions will contain terms that are quadratic in the structure constants, and it is likely that the Jacobi identity will be important in generating gravitational field configurations from their gauge theory counterparts.   Likewise, at this perturbative order we will have to deal with the 4-gluon interaction on the gauge theory side.  Presumably, as in BCJ, this can be handled by ``blowing up" the 4-gluon vertex into the product of 3-point interactions, but an explicit calculation is needed.

Despite being both on-shell and classical, the observables considered here are not simply the classical limit of tree-level scattering in quantum field theory.    This is clear due to the fact that for a fixed number of external particles, there are only a finite number of tree diagrams contributing to either the gauge theory or gravity $S$-matrix element.    On the other hand, even for a small set of initial and final particles, the classical observables we compute receive corrections from an infinite set of Feynman diagrams, with an ever-increasing number of source insertions at higher orders in an expansion in powers of $\epsilon\sim b^{3-d}\ll 1$.    As is well known, for gravitational scattering in the ultrarelativistic limit, it is possible to re-sum a subset of the diagrams at each order in the loop expansion of field theory to recover the classical limit.  See, e.g., refs.~\cite{Giddings:2011xs,Camanho:2014apa} for a review and recent discussion.   In this approach, an all orders resummation of ladder and cross ladder diagrams (scaling as distinct powers of the loop expansion parameter $(m_{Pl} b)^{2-d}\ll 1$) is needed just to reproduce the trivial $\epsilon^0$ solution in Eq.~(\ref{eq:LOmetric}).   The eikonal expansion is not as straightforward in the case of gauge theory; see~\cite{ii} for a recent effective theory formulation.    Nevertheless, by applying these resummation methods, ref.~\cite{Saotome:2012vy} argues that the massless gravitational shockwave solution can be recovered as the BCJ double copy of a subset of diagrams for massless scattering in QCD.    It may be feasible to extend this resummation to recover our radiative solutions at order $\epsilon^1$, albeit only in the special case of massless sources (in $d=4$ perturbative gravity, the sub-leading terms in the eikonal expansion have been computed in~\cite{Akhoury:2013yua}).   However, even if it were possible in principle to obtain the classical double copy relations reported here as a limit of the $S$-matrix BCJ double copy, it is obviously more transparent and less cumbersome to follow the approach pursued in this paper.

It is not surprising that the two-form gauge field $B_{\mu\nu}$, which plays a role in both the KLT relations and in the BCJ double copy of pure gauge theory, has not appeared in our analysis.    This follows from symmetry, as the simplest interaction between the particle worldline and $B_{\mu\nu}$ that respects gauge and diffeomorphism invariance is quadratic, of the form
\be
{m\over m_{Pl}^{d-2}}\int d\tau H^{\mu\nu\sigma} H_{\mu\nu\sigma}. 
\ee
Because the sources cannot couple linearly to $B_{\mu\nu}$, it can only appear in loop diagrams.   Thus at the classical level, $B_{\mu\nu}$ exchange does not contribute to the classical fields  $h_{\mu\nu}$ or $\phi$, nor is there at any order in $\hbar$ a non-vanishing $B_{\mu\nu}$ one-point function.   This situation differs in the case of spinning particles, in which case there can be a linear interaction
\be
\label{eq:linB}
{m\over m^{d/2-1}_{Pl}}\int d\tau S^{\mu\nu} v^\sigma H_{\mu\nu\sigma}.
\ee
involving the particle spin $S^{\mu\nu}(\tau)$, which must be treated as a dynamical variable in its own right (an effective Lagrangian to spinning compact objects was developed in~\cite{Porto:2005ac}).    We are currently investigating~\cite{gnridg} what sorts of gravity solutions can arise as double copies of Yang-Mills solutions with spinning color charges.   In this case, we expect to find that $B_{\mu\nu}$ will appear at the classical level, since it can be radiated by the source term in Eq.~(\ref{eq:linB}), which in turn may arise as the double copy of the chromo-magnetic dipole interaction
\be
g \int d\tau S^{\mu\nu} c^a F^a_{\mu\nu}
\ee
in the gauge theory.

Finally, as discused in sec.~\ref{sec:doublecopy}, for massless particle sources, or in the  $d\rightarrow\infty$ limit~\cite{Strominger:1981jg,BjerrumBohr:2003zd,Emparan:2013moa}, the effects of scalar exchange are systematically suppressed.   Thus in either of these limits, the gauge theory solution maps onto a solution of pure general relativity coupled to point particles.    To the extent that the finite size effects can be ignored, gluon emission from ultrarelativistic non-abelian charges double copies to gravitational radiation by interacting black holes.   Provided that the replacement patterns found in this paper persist at higher orders in powers of $G_N E/b^{d-3}\ll 1$, this observation paves the way for possible applications of BCJ double copy rules to the physics of gravitational wave sources and LIGO.    While it is not clear if there is an astrophysically significant fraction of perturbative ultrarelativistic black hole collisions\footnote{In four-dimensional general relativity, gravitational radiation from colliding ultrarelativistic black holes was first worked out in~\cite{death}, while gravitational bremsstrahlung in the post-Minkowskian expansion was formulated in~\cite{Kovacs:1977uw}.  Systematics of the large boost expansion have been discussed in~\cite{Galley:2013eba}.} that can be detected by LIGO, the efficient calculation of precision gravitational wave templates directly from Yang-Mills Feynman rules could play a role in comparisons between numerical and perturbative methods, and give insight into template models.   We leave these and other questions raised by our results for future work.

\section{Acknowledgments}

WG thanks Witek Skiba for insightful comments.   We thank Mark Wise and Chad Galley for useful discussions.  This research was partially supported by Department of Energy grants DE-FG02-92ER-40704 (WG) and DE-SC0011632 (AKR),  and by the Gordon and Betty Moore Foundation through Grant No. 776 to the Caltech Moore Center for Theoretical Cosmology and Physics (AKR).

\end{document}